# Research on financial fraud algorithm based on federal learning and big data technology

Xinye Sha*
Columbia University
New York, United States
xs2399@columbia.edu

*Abstract*—With the deepening of the digitization degree of financial business, financial fraud presents more complex and hidden characteristics, which poses a severe challenge to the risk prevention and control ability of financial institutions. At the same time, the vigorous development of big data technology provides massive potential information resources, and federated learning, as an emerging distributed machine learning paradigm, can realize multi-party data collaborative modeling under the premise of protecting data privacy. This paper firstly elaborates the basic principle, advantages and unique value of federated learning in solving data silos and protecting user privacy. Aiming at the needs of financial fraud detection, this paper discusses the design of federal learning architecture suitable for this scenario, including selecting suitable model type (such as neural network), setting reasonable data partitioning and updating rules. The central theme of the dissertation revolves around the exploration and execution of an algorithm for detecting financial fraud, which is grounded in federated learning methodologies. With a federated learning framework, each participant trains the model locally and exchanges only model parameters rather than raw data, enabling iterative optimization of the global model while protecting data privacy. To ascertain the efficacy and superiority of the suggested algorithm, a meticulous experimental investigation is both devised and executed. A real-world financial fraud dataset is selected to compare the fraud detection performance using traditional centralized learning and federated learning. Evaluation indicators include accuracy, recall rate, F1 score, etc. The findings from the experiments reveal that the federated learning-based financial fraud algorithm achieves a substantial reduction in the likelihood of data privacy breaches without compromising on high detection accuracies. Furthermore, it adeptly addresses challenges such as imbalanced data distribution and sparse sampling issues, thereby demonstrating its robust practical significance.

*Keywords—federal learning, big data, fraud recognition, neural networks*

## I. INTRODUCTION

Like a double-edged sword, the thorough digitization of financial operations has both its advantages and disadvantages. On the positive side, it significantly bolsters the ease, personalization, and worldwide accessibility of financial services, concurrently enhancing the operational efficacy of the financial ecosystem. On the other hand, it also provides a broader and secret breeding ground for financial fraud, making fraud methods more complex and changeable, and significantly more difficult to identify and prevent. Confronted with this pressing challenge, there is an urgent need for financial institutions to leverage sophisticated technological solutions in order to augment their risk management and control capabilities, thereby guaranteeing the steady and healthy functioning of financial markets. In this context, the organic combination of big data technology and federal learning provides new ideas and possibilities for solving financial fraud problems [1].

The development of big data technology has opened a window into complex economic phenomena for the financial field. Massive, high-speed and diversified data resources contain rich individual behavior patterns, market trend information and potential risk signals. Through the deep mining and intelligent analysis of these data, it can provide unprecedented depth and breadth for financial fraud detection, and improve the identification accuracy and response speed of fraud. However, the explosive growth of data does not automatically translate into the synchronous improvement of risk control ability, but exacerbates the problem of data silo to a certain extent, hinders cross-institutional and cross-industry data sharing and collaboration, and limits the effectiveness of fraud detection models [2]. In addition, with the increasingly strict data protection regulations, how to use big data to improve the efficiency of risk control while properly protecting user privacy and avoiding legal risks caused by data abuse has become a realistic issue that financial institutions must face.

Federated learning, as an emerging distributed machine learning paradigm, just provides an innovative solution to the above problems. Its core idea is to realize multi-party data joint modeling and knowledge sharing under the premise of protecting data privacy. Specifically, the participants (such as financial institutions, e-commerce platforms, telecom operators, etc.) save and process the data locally, only exchanging intermediate results such as encrypted gradients and parameters during the model training process, rather than the original data itself. The ingenious design of This avoids the risk of privacy disclosure caused by direct data exchange [3], while retaining the advantages of big data analysis, that is, to improve the generalization ability and prediction accuracy of the model by aggregating a large number of heterogeneous data.

This paper will first elaborate the basic principle of federated learning in depth, analyze how to ensure data privacy through encrypted communication, differential privacy, homomorphic encryption and other technical means, and how to achieve collaborative optimization of the model through the iterative process of local update and global aggregation [4]. Further, we will explore the unique value of federated learning in addressing data silos and facilitating cross-domain data collaboration, particularly in the specific scenario of financial fraud detection. This includes how to select a suitable model structure (such as deep neural network) according to the characteristics of fraud behavior and data characteristics, how to design a reasonable data division scheme and update rules to balance computational efficiency and model performance, and how to introduce regularization, transfer learning and other strategies to deal with problems such as uneven data distribution and sparse samples.

The core part of the thesis will introduce the design and implementation of financial fraud algorithm based on federal learning in detail. We will conduct a series of experimental studies on real-world financial fraud datasets to compare fraud detection performance using traditional centralized learning (where all data is in one place for model training) with federated learning. The evaluation indicators include but are not limited to accuracy rate, recall rate, F1 score, etc., aiming at comprehensively examining the recall rate, accuracy rate and comprehensive recognition ability of the model. It is expected that the experimental results will reveal that the fraud detection algorithm based on federated learning can significantly reduce the risk of data privacy disclosure while maintaining the high-precision identification ability, effectively deal with the practical problems such as uneven data distribution and sparse samples, and show strong practical value and promotion prospects.

Furthermore, the essay will delve into the prospective applications, existing challenges, and prospective avenues for exploration of federated learning within the realm of financial fraud prevention. We expect that this study will not only provide a new fraud detection tool for financial institutions, but also provide valuable theoretical reference and practical experience for academia and industry in the fields of data privacy protection, data sharing mechanism, distributed machine learning, etc., and jointly promote the construction and development of intelligent financial risk prevention and control system.

## II. RELATED WORK

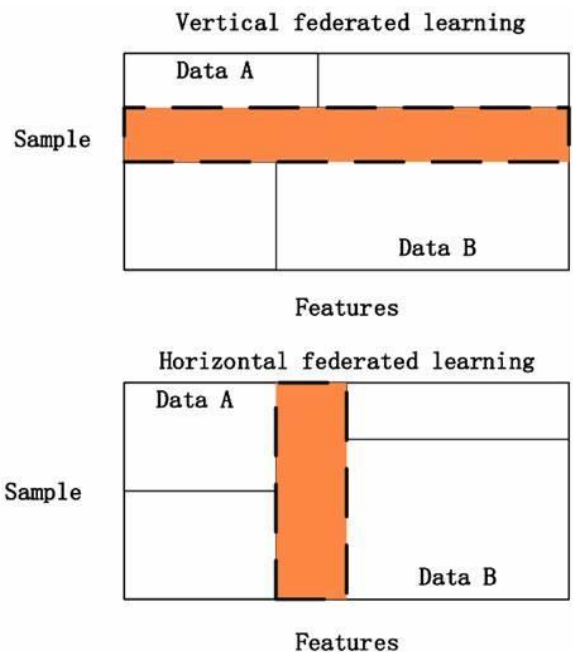

Fig. 1. Horizontal federated learning and vertical federated learning

In the application of federal learning between banks across regions, although the overlap of bank customer groups in two different cities is low, due to the similar nature of banking business, there is a large common part of the customer feature space. With this feature in mind, Google introduced a horizontal federated learning strategy in 2016 specifically for updates to Android phone models [5]. As shown in Figure 1, this horizontal federated learning framework enables individual Android users to educate their models locally using their own data sets, transferring only refined model parameters to the Android cloud infrastructure subsequently without directly sharing the raw data, significantly reducing the risk of privacy breaches. The researchers construct a secure client-server architecture, so that all clients participating in horizontal federated learning can train a global sharing model together, and the whole process ensures the data privacy is not leaked, and effectively protects the data security. Shokri and Shmatikov[6] further proposed a collaborative deep learning method, which enables each collaborator to exchange only a subset of model parameter updates on the basis of independently training the local model, further reducing the risk of data exposure. Bonawitz et al. [7] developed a secure aggregation mechanism that effectively protects the privacy of clients participating in data aggregation within their federated learning framework. In the work of Smith et al. [8], a multi-task federated learning system is presented, which enables multiple sites to learn specific tasks individually, and can use third-party data, while considering communication cost and fault tolerance, thus providing a solution with both security and efficiency. Aono et al. [9] used additive homomorphic encryption algorithm to aggregate model parameters, aiming to enhance the security protection performance of the central server and further strengthen the data security and privacy protection barrier. Ultimately, Lin et al. [10] introduced the Deep Gradient Compression technique, which markedly enhances the overall communicational efficiency of the algorithm by lessening the bandwidth necessities in large-scale distributed training scenarios. This innovation paves a fresh avenue for implementing federated learning within environments constrained by either resources or communication capacities. In summary, a series of research works have carried out in-depth exploration and technological innovation in data privacy protection, communication efficiency improvement, system security strengthening and other aspects of federated learning, providing theoretical basis and practical guidance for cross-regional banks to effectively utilize federated learning for safe and efficient cooperation modeling.

Western developed countries have shown significant advantages in the key technologies of big data distributed processing, especially in credit risk control and management. With the explosive growth of Internet data, how to efficiently deal with massive user data has become a key issue that global enterprises need to solve. In this regard, Google took the lead in proposing the classic big data batch processing technology MapReduce in 2004 [12]. Adhering to the principle of "divide and rule", the technology divides large files into multiple parts for independent processing, and then summarizes the processing results, significantly simplifying the large-scale data calculation process, and effectively reducing the communication overhead in the data transmission process. Once MapReduce was released, it quickly attracted wide attention and has now become one of the core technologies in the field of big data processing. After MapReduce, Google further launched the distributed file system GFS[13] based on its internal system operation experience. GFS provides robust support for upper-layer applications, ensures high reliability of data storage, and lays a solid foundation for large-scale data processing. Although Google upgraded the GFS system in 2013, it has not publicly released a new version of the paper. In addition to Google, other foreign technology giants have also actively explored the key technology fields of big data. Drawing on the design concept of GFS, Borthakur and Dhruba designed HDFS[14], and Lehrig et al. developed

CloudStore[15]. In view of the poor performance of GFS in processing images and small file transmission, Facebook launched Haystack[16] system which is suitable for large amounts of small data processing. By implementing multiple logical files to share the same physical file and using caching technology, Haystack successfully overcomes the aforementioned defects and significantly improves the processing efficiency of small files.

## III. FEDERATED LEARNING

### A. concept

Federated learning is a distributed learning approach to machine learning that aims to train models without the need to centralize all the data into one place. In traditional centralized learning, all data is collected on a central server for training. In contrast, federated learning distributes the training process of the model to various local devices or data owners. This means that raw data does not need to leave the device or data center, only the parameters of the model are sent to a central server for aggregation.

Federated learning usually consists of the following steps:

Choose Model Schema: Determine the model type and schema to use in federated learning. Select Participants: Identify the device, terminal, or data owner participating in federated learning. Initialize the model: Initializes the model among the participants. Local training: Each participant trains the model using local data. Parameter aggregation: Aggregate model parameters, usually by averaging or other methods to combine parameter updates from various participants. Update Global model: Apply aggregated parameter updates to the global model. Iterative training: Repeat the above steps until the training termination condition is met.

The benefits of federated learning include data privacy, as raw data does not need to be shared; Reduce data transfer requirements because only model parameters need to be transferred; As well as facilitating the personalization of the model, as each participant can be trained locally based on their local data.

### B. Life cycle

The typical workflow of federated learning is often driven by algorithm engineers who develop models for specific applications. For example, experts in the field of natural language processing can develop next word prediction models for virtual keyboards. As shown in Figure 2, federated learning consists of the following steps:

1. Problem determination: Model developers specify the problem to be solved using federated learning.

2. Client configuration: Clients (such as mobile apps) are deployed and configured to collect data sets required for local model training. An application often retains specific data, such as an SMS application archiving text messages or a photo management application storing photographs. Occasionally, these applications might also necessitate preserving supplementary information, like records of user interactions, to facilitate labeling functionalities.

3. Simulation prototype: Model developers use auxiliary data sets to prototype the architecture of the model in a federated learning simulation environment and test the learning model hyperparameters.

4. Concurrent Model Training: Undertake multiple federated learning tasks concurrently to educate an array of model architectures or models that have been fine-tuned with distinct hyperparameters.

5. Model evaluation: After the model is fully trained, the model is analyzed and evaluated, and the better model is selected. The evaluation may involve calculating metrics on standard datasets in the data center, or conducting a joint evaluation where the model is pushed to constrained clients for evaluation with their local private datasets.

6. Deployment: After selecting the model, proceed to the standard model deployment process.

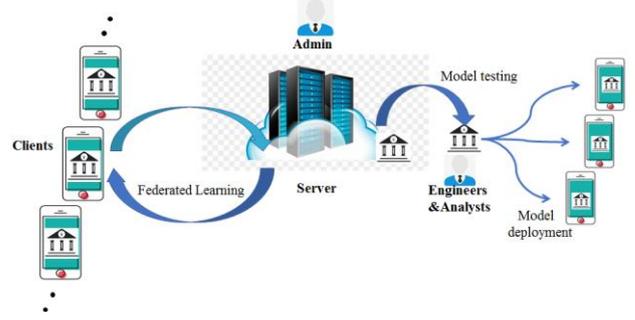

Fig. 2. The lifecycle of the model in a federated learning system

## IV. FEDERAL AVERAGE ALGORITHM

Suppose a group of K financial institutions are trained in a federal learning model, each with its own independent local private data set containing sensitive user information:

$$R_s = \{m_i^s, n_i^s\} (s=1,2,\ldots,S). \quad (1)$$

Where, $m_i^s$ is the feature vector in the data sample, $n_i^s$ is the label of the sample, and $n_i$ represents the data set size of the I-th institution node participating in federated training. The objectives of the central joint model training are:

$$\min_{t \in \mathbb{R}^d} l(m,n;t) \quad (2)$$

In formula (2), l (m, n; t) represents the total number of data set samples of all mechanism nodes, that is, minimizing the loss value of the model on all data involved in training. According to the distribution of data set size in each local institution, the objective function can be expressed as:

$$l(m,n;t) = \sum_{s=1}^{S} \frac{n_k}{n} L_k(m_k,n_k;t) \quad (3)$$

Where $L_k$ represents the loss function of local training of each institution.

$$L_k(m_k,n_k;t) = \frac{1}{n_k} \sum_{i \in D_k} l(m_i,n_i;t) \quad (4)$$

In the case that all clients participate in training and the local model is updated by random gradient descent (SGD), the gradient of local model training update for the t round is:

$$\nabla l = \sum_{s=1}^{S} \frac{n_k}{n} g_k \quad (5)$$

## V. Fraud detection model construction

Fraud detection aims to accurately distinguish between fraudulent and non-fraudulent samples in the data set, i.e. the binary classification task of supervised learning. In fraud detection, there are many machine learning algorithms to choose from. The author previously took part in the research and the development of an XGBoost based system for credit card fraud detection[17].. Considering the effect of fraud detection and the fusion of algorithm and federated learning architecture, this paper chooses multi-layer perceptron (MLP) as the fraud detection model. A Multilayer Perceptron (MLP) represents a category of Artificial Neural Networks (ANNs) characterized by a composition comprising an input layer, one or more hidden layers, and an output layer. Notably, while both the input and output layers are designed with a singleton node each, the hidden layer is capable of accommodating multiple nodes. Each layer is made up of many identical neural units that are connected to each other for parallel data processing, enabling powerful information processing capabilities. The node data in the neural network is processed by input, linear and nonlinear transformation, output, etc., and multi-layer nodes interact with each other to form a huge data processing network. This kind of network can deal with complex mapping relationships and is suitable for learning complex relationships and patterns in the real world. In the financial fraud detection scenario, MLP can help find the law of fraud cases and analyze the correlation between transaction data and fraud determination.

MLP (Multilayer Perceptron) is a common type of Artificial Neural Network (ANN). It is comprised of a multi-tiered architecture involving numerous neurons, typically encompassing an input layer, one or more intermediary or hidden layers, and ultimately, an output layer.

1. Input Layer: The input layer receives the features from the data set and passes them to the next layer of the neural network. Each input layer node corresponds to a feature of the data.

2. Hidden Layers: Hidden layers are between the input layer and the output layer, and are responsible for a series of nonlinear transformations and feature extraction of input data. Each hidden layer consists of multiple neurons (nodes), each of which receives input from the previous layer and processes it through activation functions before passing it to the next layer.

3. The Output Layer: This layer absorbs the signals transmitted by the terminal hidden layer and produces the ultimate output. In the context of classification tasks, the output layer customarily employs softmax activation functions to render probability distributions across all classes. Conversely, when tackling regression problems, the configuration of the output layer is simplified to a single node, tasked with emitting the predicted numerical value directly.

The customary training methodology for MLPs revolves around the Backpropagation algorithm. This algorithm modifies model parameters through the calculation of a loss function, with the primary objective being to minimize the discrepancy between predictions and actual outcomes. Throughout the training phase, the weights and biases of the model undergo adjustments, facilitated by optimization techniques such as gradient descent. These adjustments ensure that the model progressively aligns itself more accurately with the training dataset.

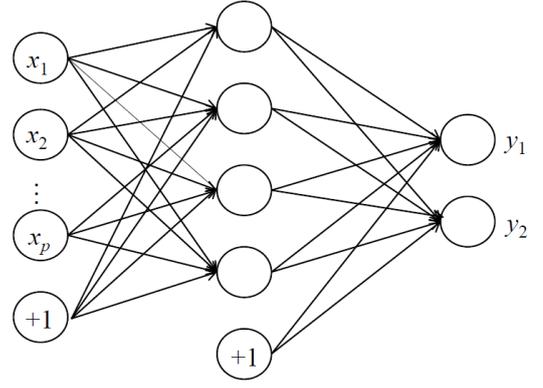

Fig. 3. Structure of MLP

Each node in the MLP neural network is a perceptron, which simulates the function of neurons in the biological neural network. The feature value output from the upper layer is input into the neuron, and after the feature transformation within the neuron, including linear weighting and nonlinear function activation processing, as shown in Figure 4, the result is output to the neuron of the next layer.

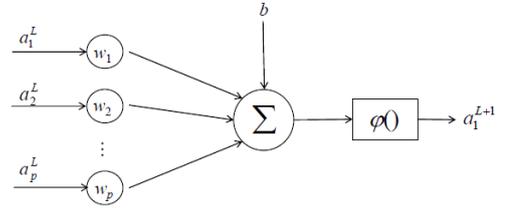

Fig. 4. Internal feature transformation operation of MLP neurons

## VI. Experiment

### A. Data set

This paper utilizes an experimental dataset sourced from the ULB Machine Learning Group, which comprises transaction records from European credit card holders. The dataset encompasses a total of 284,807 entries, among which merely 492 instances represent fraudulent transactions, constituting a mere 0.17% of the entire dataset. This indicates a stark imbalance with a normal-to-fraudulent sample ratio of 578:1. In safeguarding users' privacy, the dataset's original 30 features have been transformed through Principal Component Analysis (PCA).

### B. Evaluation index

Accuracy, precision, and recall are common metrics used to evaluate the performance of classification models based on the confusion Matrix, a table that describes the relationship between the model's predictions and the real label, as shown in Table I.

TABLE I. Confusion Matrix

| Truth / forecast | Cheat sample | Normal sample |
|---|---|---|
| Cheat sample | True Class (TP) | False positive class (FP) |
| Normal sample | Pseudonegative class (FN) | True negative class (TN) |

Accuracy provides a global view of the model's ability to make correct predictions as a whole, as shown in formula (6):

$$\text{Accuracy} = \frac{TP + TN}{TP + FP + TN + FN} \quad (6)$$

Accuracy assesses the fraction of instances that the model correctly identifies as positive among those it deems positive. Alternatively stated, it reflects the ratio of true positive identifications (TP) to the sum of all instances classified as positive (TP plus FP). The formula for its computation is thus formulated as follows:

$$\text{Precision} = \frac{TP}{TP + FP} \quad (7)$$

Recall, or sensitivity, gauges the model's efficacy in detecting every actual positive instance, representing the fraction of true positives correctly flagged by the model relative to all actual positives in the dataset (TP plus FN). The mathematical expression for its calculation is provided below:

$$\text{Recall} = \frac{TP}{TP + FN} \quad (8)$$

F1 Score is the harmonic average of accuracy and recall, reaching a maximum when accuracy and recall are equal. Formula 1 Score is calculated as follows:

$$\text{F1 Score} = 2 \times \frac{\text{Precision} \times \text{Recall}}{\text{Precision} + \text{Recall}} \quad (9)$$

*C. Contrast algorithm*

LR (Logistic Regression) model [1] : In simple terms, LR first maps the data linearly and then converts the result to a value between 0 and 1 through a logical function.

DT (Decesion Trees) model [2] : The method classifies the samples through a series of rules. First, one of the features of the sample data is selected as the basis for splitting the current node. Then, the current node is constantly split through a recursive way until all the samples of the current node belong to the same category.

*D. Experimental result*

TABLE II. EXPERIMENTAL RESULTS OF DIFFERENT MODELS UNDER FOUR INDEXES

| Model | AUC | PR | RE | F1 |
|---|---|---|---|---|
| LR | 0.75 | 0.82 | 0.62 | 0.71 |
| DT | 0.68 | 0.85 | 0.57 | 0.70 |
| OUR | 0.82 | 0.89 | 0.68 | 0.77 |

Table II shows the experimental results of different models under the four indicators. It can be seen that the AUC value of logistic regression model is 0.75, indicating that its ability to distinguish positive and negative samples under different thresholds is strong but not as good as that of the other two models. The average Precision of the model reaches 0.82, which shows that it has high prediction accuracy for positive samples and few false positives. The recall rate is 0.62, which shows that it is not good in preventing omission. The F1 score of 0.71 reflects the model's balance between accuracy and recall, with overall performance being the lowest of the three. The AUC value of the decision tree model is 0.68, indicating that it has the smallest area under the ROC curve and the weakest ability to distinguish positive and negative samples. Although its average Precision is 0.85, demonstrating high reliability for positive predictions, its recall rate is only 0.57, indicating a significant shortfall in preventing underreporting. The F1 score of 0.70 indicates that the model achieves a slightly lower balance between accuracy and recall than the logistic regression model, and the overall performance is in the middle. Our model has shown excellent performance in every index. Its AUC value is as high as 0.82, which shows strong ability to distinguish positive and negative samples and generalization ability. The average Precision of 0.89 means that the prediction accuracy of the positive sample is extremely high and there are very few false positives. The recall rate is 0.68, which is not as prominent as the accuracy rate, but better than the logistic regression model in terms of recall ability. The F1 score reached 0.77, which reflects that the model has achieved a good balance between accuracy and recall rate, and the overall performance is optimal.

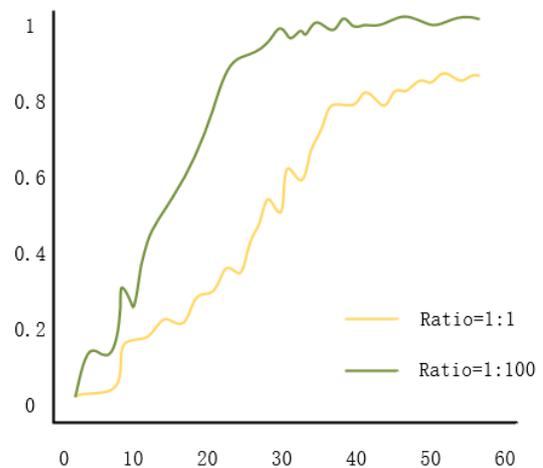

Fig. 5. Sensitive test of sampling ratio of fraud and legitimate transactions

Figure 5 shows the variation of west energy under AUC index with different sampling rates. The horizontal axis represents the number of samples and the vertical axis represents the AUC value. The two curves represent two different sampling rates: the yellow line represents the sample rate of 1:1, and the green line represents the sample rate of 1:100. As can be seen from the figure, with the increase of the number of samples, the AUC values under the two sampling rates show an increasing trend. When the number of samples is small, the curve with a sampling rate of 1:1 (yellow) is higher than that with a sampling rate of 1:100 (green), indicating that a sampling rate of 1:1 can obtain a higher AUC value when the number of samples is small. However, when the number of samples increases to a certain extent, the curve with a sample rate of 1:100 (green) begins to exceed the curve with a sample rate of 1:1 (yellow), indicating that a sample rate of 1:100 can obtain a higher AUC value in the case of a large number of samples.

VII. CONCLUSION

Faced with the increasingly complex and hidden fraud in the process of digitization of financial business, financial institutions are faced with a severe test of risk prevention and control. In this context, the rich resources of big data technology and emerging federated learning technologies provide new ways to address the challenges. Federated learning takes data privacy protection as its core, effectively breaks data silos through distributed collaborative modeling, and highlights its unique value in the financial field. This paper deeply analyzes the principle, advantages and contributions of federal learning in privacy protection, and specifically designs a federal learning architecture for financial fraud detection, which covers key design links such as model type selection, data partitioning and update rules. The core of the research is a financial fraud algorithm based on federal learning, which not only ensures data privacy and security, but also realizes efficient global model optimization by locally training the model and only exchanging parameters. In order to verify the actual efficiency and advantages of the algorithm, this paper conducted rigorous experimental research, selected real financial fraud data sets, compared the performance of centralized learning and federal learning in fraud detection, measured by multidimensional indicators such as accuracy rate, recall rate and F1 score. The experimental results show that the financial fraud algorithm based on federal learning not only maintains a high detection accuracy, but also significantly reduces the risk of data privacy disclosure, and can effectively deal with practical problems such as uneven data distribution and sparse samples, which fully demonstrates its strong practicability and technical advantages in financial anti-fraud practice. Therefore, the introduction of federal learning technology to build a financial fraud detection system under privacy protection has important value and broad application prospects for improving risk prevention and control.